\documentclass[twocolumn,PRB,showpacs,preprintnumbers,amsmath,amssymb]{revtex4}
\usepackage{graphicx}% Include figure files
\usepackage{dcolumn}% Align table columns on decimal point
\usepackage{bm}% bold math

\begin{document}

\title{Narrow superconducting window in Ni-doped LaFeAsO}

 \author{Guanghan Cao$^{1}$}
 \email{ghcao@zju.edu.cn}

 \author{Shuai Jiang,$^{1}$ Xiao Lin,$^{1}$ Cao Wang,$^{1}$ Yuke Li,$^{1}$ Zhi Ren,$^{1}$ Qian Tao,$^{1}$ Chunmu Feng,$^{2}$ Jianhui Dai$^{1}$}
 \author{Zhu-an Xu$^{1}$}%
 \email{zhuan@zju.edu.cn}
 \author{Fu-Chun Zhang$^{1,3}$}
\affiliation{$^{1}$Department of Physics, Zhejiang University, Hangzhou 310027, China\\
$^{2}$Test and Analysis Center, Zhejiang University, Hangzhou
310027, China\\ $^{3}$Department of Physics, the University of Hong
Kong, Hong Kong, China}
\date{\today}% It is always \today, today,

\begin{abstract}
We have studied Ni-substitution effect in LaFe$_{1-x}$Ni$_{x}$AsO
($0\leq x \leq0.1$) by the measurements of x-ray diffraction,
electrical resistivity, magnetic susceptibility, and heat capacity.
The nickel doping drastically suppresses the resistivity anomaly
associated with spin-density-wave ordering in the parent compound.
Superconductivity emerges in a narrow region of $0.03\leq x
\leq0.06$ with the maximum $T_c$ of 6.5 K at $x$=0.04, where
enhanced magnetic susceptibility shows up. The upper critical field
at zero temperature is estimated to exceed the Pauli paramagnetic
limit. The much lowered $T_c$ in comparison with
LaFeAsO$_{1-x}$F$_{x}$ system is discussed.
\end{abstract}

%74.70.Dd    Ternary, quaternary, and multinary compounds (including Chevrel phases, borocarbides, etc.)
%74.62.Dh    Effects of crystal defects, doping and substitution
%74.25.Dw    Superconductivity phase diagrams
%74.25.Ha    Magnetic properties
%74.10.+v    Occurrence, potential candidates

\maketitle

\section{\label{sec:level1}Introduction}

The discovery of superconductivity at 26 K in
LaFeAsO$_{1-x}$F$_{x}$\cite{Kamihara08} and the subsequent findings
of the enhanced superconductivity with $T_c$ up to 56
K\cite{Sm,Ce,55K,56K} in a series of iron-arsenides have stimulated
enormous research interest. It has been suggested that both
electronic correlations and multi-orbital/band effects should play
important roles\cite{Mazin,Boeri,DaiX,ZhangSC,Lee&Wen,Si,Kuroki}.
The prototype parent compound, LaFeAsO, undergoes a structural phase
transition at 155 K \cite{Dai,Nomura}, followed by a collinear
antiferromagnetic (AFM) spin-density-wave (SDW) transition at lower
temperature\cite{wnl,Dai}. Electron/hole doping into the FeAs layers
suppresses the long-range SDW order, in favor of
superconductivity\cite{wnl,wen}. This phenomenon is apparently
analogous to that in cuprates, where superconductivity is induced by
doping of charge carriers into an AFM Mott insulator. The
iron-arsenides, however, also show remarkable differences from the
cuprates. For example, the parent compounds of iron-arsenides show
itinerant character of Fe-3$d$
electrons\cite{Singh,LuZY,Tesanovic,Wu}, while Cu-3$d$ electrons in
the parent compounds of the cuprates are localized.

As inferred in the band structure calculations for LaO$M$As ($M$ =
Mn, Fe, Co and Ni)\cite{Xu}, \emph{Fe-site doping } with Co or Ni in
the parent compounds of iron-arsenides may also introduce additional
electrons, hence possibly inducing superconductivity. This has been
experimentally realized for the Co doping with $T_c\sim 13$ K in
LaFe$_{1-x}$Co$_x$AsO\cite{Sefat1,WangC} and $T_c= 22$ K in
BaFe$_{1.8}$Co$_{0.2}$As$_2$\cite{Sefat2}. Since Ni atoms have one
more electron than Co, one would expect that substitution of Fe with
Ni introduces carriers more effectively. A possible hint comes from
the Ni-based arsenide analogue, LaNiAsO, which is a superconductor
with $T_c \sim$ 2.5 K \cite{Luo,jssc}. The normal state of the
LaNiAsO superconductor is Pauli-paramagnetic\cite{jssc}, suggesting
that the Ni-3$d$ electrons have an itinerant character.

In this paper we report the realization of superconductivity in
LaFe$_{1-x}$Ni$_{x}$AsO ($0\leq x \leq0.1$). Superconductivity has
been observed in a narrow region of $0.03\leq x \leq0.06$ with a
lowered maximum $T_c$ of 6.5 K. The optimal doping level is found to
be about half of that in LaFe$_{1-x}$Co$_{x}$AsO (Ref.~\cite{WangC})
system. The occurrence of superconductivity by Ni doping at Fe-site
contrasts sharply with severe suppression of superconductivity by
the Cu-site doping with Ni in cuprate superconductors.

\section{\label{sec:level2}Experimental}

Polycrystalline LaFe$_{1-x}$Ni$_{x}$AsO samples were synthesized by
solid state reaction in vacuum, similar to previous
report\cite{WangC}. Powders of LaAs, La$_{2}$O$_{3}$, FeAs,
Fe$_{2}$As and NiO were weighed according to the stoichiometric
ratios of LaFe$_{1-x}$Ni$_{x}$AsO ($x$=0, 0.01, 0.02, 0.03, 0.04,
0.05, 0.06, 0.08 and 0.1), and thoroughly mixed in an agate mortar
and pressed into pellets under a pressure of 2000 kg/cm$^{2}$. The
pellets were sealed in evacuated quartz tubes, then heated uniformly
at 1433 K for 48 h, and finally cooled by shutting off the furnace.

The resultant samples were characterized by powder x-ray diffraction
(XRD) with Cu K$_{\alpha}$ radiation. The XRD diffractometer system
was calibrated using standard Si powders. The detailed structural
parameters were obtained by Rietveld refinements, using the
step-scan XRD data with $10^{\circ}\leq 2\theta \leq 120^{\circ}$.
The typical $R$ values of the refinements are: $R_{\text{F}}\sim$
2.8\%, $R_{\text{I}}\sim$ 4.6\%, and $R_{\text{wp}}\sim$ 13\%. The
goodness-of-fit parameter, $S$=${R_{\text{wp}}/R_\text{exp}}\sim$
1.6, indicating good reliability of the refinement.

The electrical resistivity was measured with a standard
four-terminal method. Samples were cut into a thin bar with typical
size of 4$\times$2$\times$0.5 mm$^3$. Gold wires were attached onto
the samples' abraded surface with silver paint. The size of the
contact pads leads to a total uncertainty in the absolute values of
resistivity of ¡À10 \%. The measurements of magnetoresistance and
heat capacity were carried out on a Quantum Design physical property
measurement system (PPMS-9). Temperature dependence of magnetization
was measured on a Quantum Design magnetic property measurement
system (MPMS). In the measurements of normal state susceptibility,
the background data from the sample holder were removed. For the
measurement of the superconducting (SC) transitions, both the
zero-field cooling (ZFC) and field cooling (FC) protocols were
employed under the field of 10 Oe.

\section{\label{sec:level3}Results and discussion}

Figure 1(a) shows XRD patterns of the representative samples of
LaFe$_{1-x}$Ni$_{x}$AsO. The XRD peaks are well indexed based on a
tetragonal cell with the space group of P$4/nmm$, indicating that
the samples are essentially single phase. The lattice parameters are
plotted in the inset as functions of $x$. With the increase of Ni
doping the $a$ axis increases slightly, while the $c$ axis shrinks
remarkably. The cell volume is consequently decreased by the
incorporation of Ni.

\begin{figure}
\includegraphics[width=7cm]{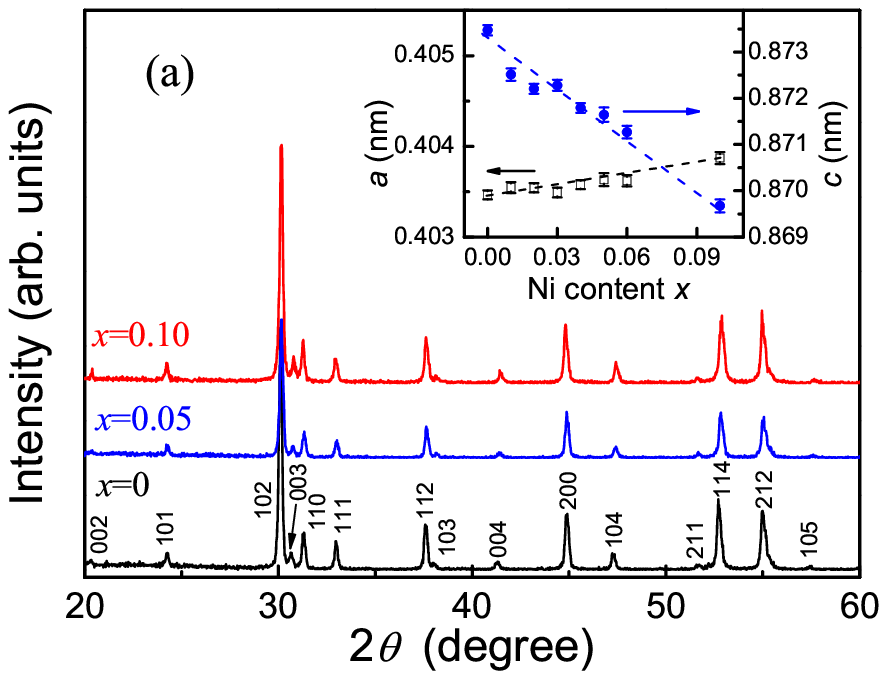}
\includegraphics[width=7cm]{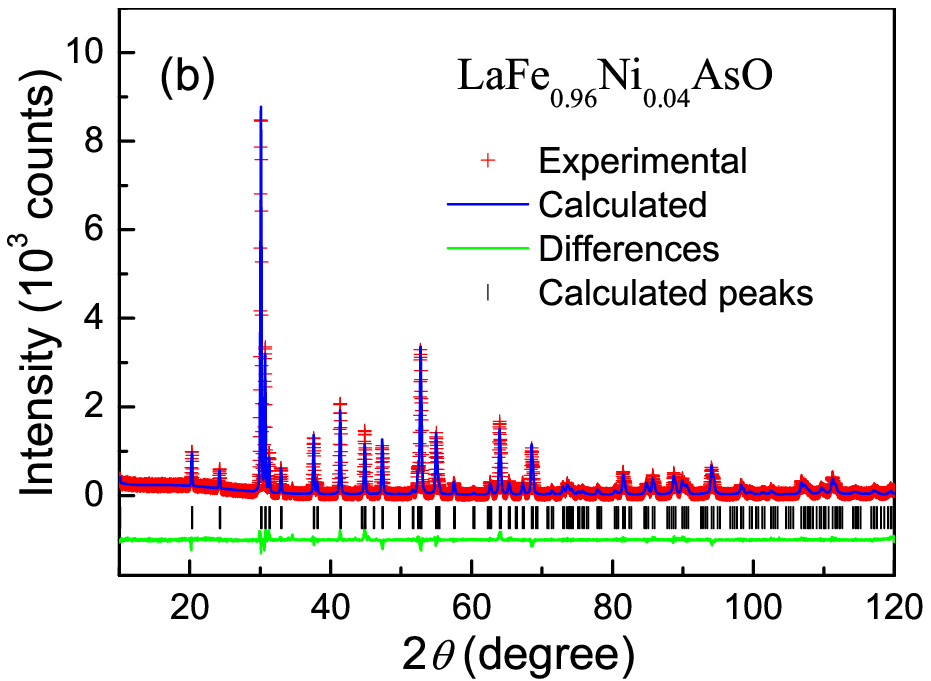}
\caption{(Color online) (a) Powder x-ray diffraction patterns of
representative samples of LaFe$_{1-x}$Ni$_{x}$AsO. The inset shows
the lattice constants as functions of Ni content. (b) An example of
Rietveld refinement profile for $x$=0.04.}
\end{figure}

The crystallographic parameters were obtained by the Rietveld
refinement [Fig. 1(b)] based on ZrCuSiAs-type structure. Table
~\ref{tab:table1} compares the structural data of undoped and
Ni-doped (by 4 at.\%) samples. The Ni doping enlarges the Fe-Fe
spacing slightly, but compresses the FeAs layers significantly. In
other word, the most remarkable effect of Ni doping on the crystal
structure is that As atoms are pulled towards the Fe planes.

\begin{table}
\caption{\label{tab:table1}Crystallographic data of
LaFe$_{1-x}$Ni$_{x}$AsO ($x$=0 and 0.04) at room temperature. The
space group is P$4/nmm$. The atomic coordinates are as follows: La
(0.25,0.25,$z$); Fe/Ni (0.75,0.25,0.5); As (0.25,0.25,$z$); O
(0.75,0.25,0).}
\begin{ruledtabular}
\begin{tabular}{lcr}
Compounds&LaFeAsO&LaFe$_{0.96}$Ni$_{0.04}$AsO\\
\hline
$a$ ({\AA}) & 4.0357(3) &4.0376(3)\\
$c$ ({\AA}) & 8.7378(6) &8.7208(6)\\
$V$ ({\AA}$^3$) & 142.31(2) & 142.17(2)\\
$z$ of La & 0.1411(2)& 0.1422(2)\\
$z$ of As & 0.6513(3) & 0.6505(3)\\
FeAs-layer thickness ({\AA}) & 2.644(2) &2.624(2)\\
Fe-Fe spacing ({\AA}) & 2.8536(3) &2.8550(3)\\
As-Fe-As angle ($^{\circ}$) & 113.5(1) &114.0(1)\\
\end{tabular}
\end{ruledtabular}
\end{table}

Figure 2 shows temperature dependence of resistivity ($\rho$) in
LaFe$_{1-x}$Ni$_{x}$AsO. The parent compound shows a resistivity
anomaly below 155 K. This resistivity anomaly has been identified as
due to a structural phase transition associated with SDW
instability.\cite{Dai,Nomura,wnl,Oak} Upon doping with 1\% and 2\%
Ni, the anomaly temperature $T_{\text{anom}}$ is suppressed to 105 K
and 75 K, respectively. As $x$ increases to 0.03, a tiny anomaly in
$\rho$ can be detectable at 50 K, meanwhile the resistivity drops to
zero below 5.5 K, suggesting emergence of superconductivity. The SC
transition temperatures are 6.5 K and 3.4 K for the samples of
$x$=0.04 and 0.05, respectively, as shown in the inset of Fig. 2.

\begin{figure}
\includegraphics[width=7cm]{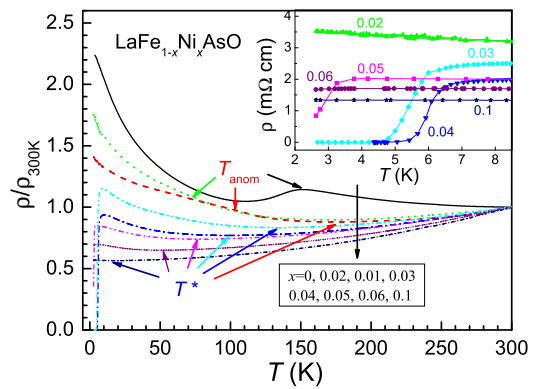}
\caption{(Color online) Temperature dependence of resistivity of
LaFe$_{1-x}$Ni$_{x}$AsO samples. The data are normalized to
$\rho_{\text{300K}}$. The temperatures of resistivity peaks/humps
($T_{\text{anom}}$) and minima ($T^*$) are respectively marked. The
inset shows an expanded plot.}
\end{figure}

Figure 3 shows SC diamagnetic transitions in
LaFe$_{1-x}$Ni$_{x}$AsO. Although samples with $x\leq $ 0.02 show no
diamagnetic signal above 1.8 K, magnetic expelling/screening can be
clearly seen for $0.03\leq x\leq $ 0.06 at low temperatures. The
magnetic shielding fraction of the sample of $x$=0.04 are estimated
to be 45\%, confirming bulk superconductivity. The diamagnetic curve
shows step-like feature, probably due to sample inhomogeneity and/or
an intergrain SC transition. The diamagnetic signal for other SC
samples is much lower, implying that the SC region would be even
narrower if samples were homogeneous. Similar phenomena were also
observed in LaFe$_{1-x}$Co$_{x}$AsO systems.\cite{Sefat1,WangC}

\begin{figure}
\includegraphics[width=7cm]{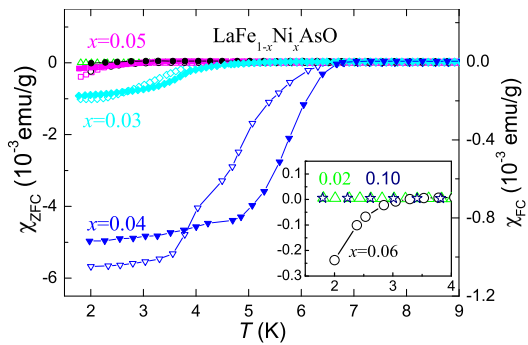}
\caption{(Color online) Temperature dependence of magnetic
susceptibility in LaFe$_{1-x}$Ni$_{x}$AsO. Note that the open and
filled symbols denote ZFC and FC data, respectively. The inset is an
expanded plot for showing the data of $x$=0.02, 0.06 and 0.10.}
\end{figure}

To verify bulk superconductivity further, we performed specific-heat
($C$) measurement for the sample of $x$=0.04. The result is shown in
Fig. 4. A specific heat anomaly can be seen at $T_{\text{c}} \sim
6.5$ K (a tiny anomaly at about 8 K might be related to the trace
residual SDW transition). In the temperature range from 6.7 to 10 K,
the specific heat can be well described by the sum of electronic and
lattice contributions: $C=\gamma T+\beta T^{3}$. Therefore, the
linear fit for $C/T$ versus $T^2$ gives the electronic specific-heat
coefficient $\gamma$=5.74 mJ/(mol$\cdot$K$^2$) and the lattice
specific-heat coefficient $\beta$=0.254 mJ/(mol$\cdot$K$^4$). The
Debye temperature $\theta_{\text{D}}$ is then calculated to be 285
K, using the formula $\theta_{\text{D}}$=$(12\pi RN/5\beta)^{1/3}$,
where $N$ = 4 and $R$ = 8.314 J/(mol$\cdot$K). The value of
$\theta_{\text{D}}$ is close to that of LaFeAsO (282 K, Ref.
~\cite{wnl}) and that of LaFeAsO$_{0.89}$F$_{0.11}$ (308 K,
Ref.~\cite{Kohama}). The value of $\gamma$ is also comparable to
those of LaFeAsO$_{1-x}$F$_{x}$ samples (Ref.~\cite{wnl,Kohama}).

\begin{figure}
\includegraphics[width=7cm]{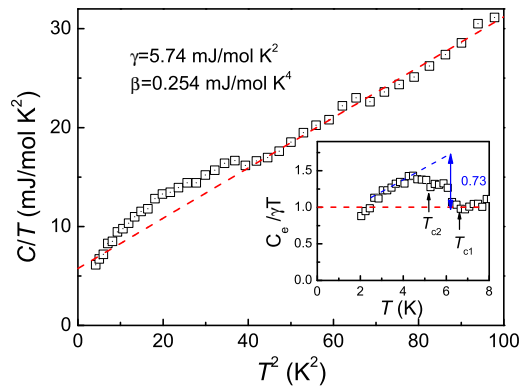}
\caption{(Color online) Curve of $C/T$ versus $T^2$ for the sample
of $x$=0.04 in LaFe$_{1-x}$Ni$_{x}$AsO system under zero field. The
inset shows $C_{\text{e}}/\gamma T$ as a function of temperature,
where $C_\text{e}$ denotes electronic specific heat.}
\end{figure}

After subtracting the lattice contribution to the specific heat, the
specific-heat jump at the SC transition can be obviously seen,
confirming bulk superconductivity. The the dimensionless parameter
$\Delta C_{\text{e}}/\gamma T$ at $T_c$ is estimated to be 0.73,
much lower than the expected value of 1.43 for an isotropic SC gap.
This observation is similar to that in
LaFeAsO$_{1-x}$F$_{x}$.\cite{Kohama} One may also see two
transitions at 5 and 6.5 K, in accordance with the above step-like
diamagnetic susceptibility curve. This phenomenon is probably due to
the sample inhomogeneity and/or intergrain SC transition as
mentioned above, however, the possibility of multiband
superconductivity cannot be fully ruled out.

Fig. 5 shows suppression of SC transition in resistivity under
magnetic fields for the sample of $x$=0.04. The applied field shifts
the SC transition towards lower temperatures, and the transition
becomes broadened. The inset plots the temperature dependence of
$T_{c}$(\emph{H}), defined as the temperature where the resistivity
falls to one half of the normal-state value. The initial slope
$\mu_{0}$$\partial$$H_{c2}$/$\partial$$T$ near $T_{c}$ is $-$3.81
T/K, which leads to an estimated upper critical field at zero field
$\mu_{0}H_{c2}$(0) $\sim$ 17 T using WWH model (Ref.~\cite{WHH}).
This value of upper critical field exceeds the Pauli paramagnetic
limit $\mu_{0}H_{\text{P}}=1.84 T_{c} \sim$ 12 T (Ref.~\cite{Hp}).
Similar observations have been reported in LaFeAsO$_{1-x}$F$_{x}$
system.\cite{Hunte,Fuchs}. The high critical field makes
LaFe$_{0.96}$Ni$_{0.04}$AsO fundamentally different from the LaFeNiO
superconductor (Ref.~\cite{Luo,jssc}).

\begin{figure}
\includegraphics[width=8cm]{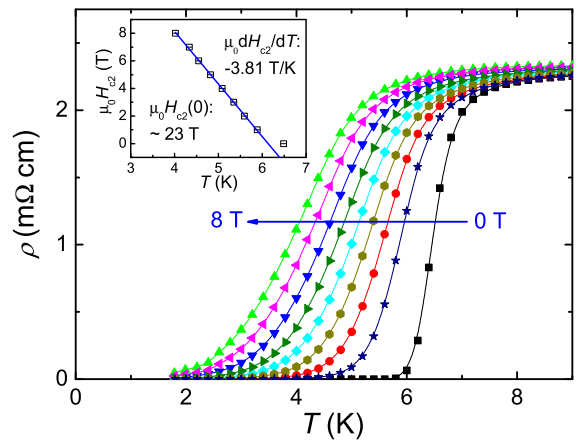}
\caption{(Color online) SC transitions in resistivity under magnetic
fields ($\mu_{0}H$=0, 1, 2, 3, 4, 5, 6, 7 and 8 T) for
LaFe$_{0.96}$Ni$_{0.04}$AsO sample. The inset shows the upper
critical fields as a function of temperature.}
\end{figure}

The normal-state susceptibility $\chi$ of the Ni-doped samples is
shown in Fig. 6. The $\chi(T)$ are characterized by linear decrease
at high temperatures as well as Curie-like upturn below $\sim$ 100
K. The Curie-like upturn was found to be sensitive to sample's
quality. Generally, better sample shows smaller susceptibility
upturn. Thus, the susceptibility upturn is mainly due to an
extrinsic origin (such as defects and trace impurities). The linear
$T$-dependence of $\chi$ was experimentally demonstrated in
LaFeAsO$_{1-x}$F$_{x}$\cite{Klingeler} and
BaFe$_2$As$_2$\cite{WangXF} systems, and was discussed in terms of
the "preformed SDW moments" \cite{GMZhang}. The measured $\chi(T)$
can thus be fitted with the formula,
\begin{equation}
\chi(T)=\chi_{0}+\alpha T+\frac{C}{T},\label{sus}
\end{equation}
where the $T$-independent term $\chi_0$ contains Pauli paramagnetic
susceptibility ($\chi_\text{P}$)\cite{note} from itinerant electrons
and Larmor diamagnetic susceptibility ($\chi_\text{core}$) from
ionic cores. It is noted that the change in $\chi_0$ with the
Ni-doping is primarily due to the variation of $\chi_\text{P}$.

\begin{figure}
\includegraphics[width=7cm]{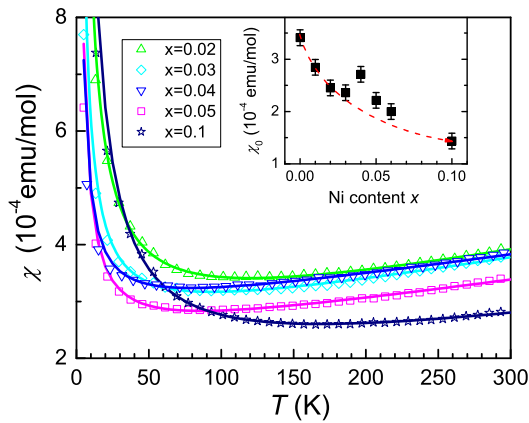}
\caption{(Color online) Temperature dependence of magnetic
susceptibility in LaFe$_{1-x}$Ni$_{x}$AsO. The applied field is 1000
Oe. The solid lines are fitted result using Eq. (1). The inset plots
the $T$-independent term $\chi_{0}$ as a function of Ni content. The
dashed line is a guide to the eye.}
\end{figure}

To avoid the influence of the structural transition for $0\leq x\leq
$ 0.02, we made the fitting using different range of data as
follows: 160 K $\leq T\leq$ 300 K for $x$=0, 110 K $\leq T\leq$ 300
K for $x=$0.01, 75 K $\leq T\leq$ 300 K for $x=$0.02, 50 K $\leq
T\leq$ 300 K for $x=$0.03, and 30 K $\leq T\leq$ 300 K for $x>$0.03.
The fitted parameters are listed in Table ~\ref{tab:table2}. It is
shown that the $\alpha$ values are about $4\times10^{-7}$
emu$\cdot$mol$^{-1}$K$^{-1}$, almost independent of Ni-doping $x$,
similar to the case reported in
LaFeAsO$_{1-x}$F$_{x}$\cite{Klingeler}. The extracted $\chi_0$,
shown in the inset of Fig. 6, tends to decrease with increasing
Ni-doping, with an enhancement in the SC regions centered at
$x$=0.04. The decrease of $\chi_{0}$ with $x$ reflects the electron
doping since band calculations show a negative d$N(E)$/d$E$ at Fermi
level.\cite{Singh,Xu} The extra susceptibility in the SC regime
resembles the behavior of thermopower in SmFe$_{1-x}$Co$_{x}$AsO
system,\cite{WangC} implying the importance of spin fluctuations for
the superconductivity. Besides, the enhanced spin fluctuations are
also evidenced by the relatively high value of Wilson ratio, defined
as $R_{\text{W}}=\frac{\pi^{2}
k_{\text{B}}^{2}\chi_{\text{P}}}{3\gamma \mu_{\text{B}}^{2}}$. Since
the $\chi_\text{core}$ of LaFe$_{1-x}$Ni$_{x}$AsO is about
$-1.0\times 10^{-4}$ emu$\cdot$mol$^{-1}$,\cite{core}
$\chi_{\text{P}}$ is thus about 3.7$\times 10^{-4}$
emu$\cdot$mol$^{-1}$ for $x$=0.04, giving $R_{\text{W}}$ of 4.7.

\begin{table}
\caption{\label{tab:table2}Fitted parameters using Eq. (1) for
LaFe$_{1-x}$Ni$_{x}$AsO system. The units of $\chi_{0}$, $\alpha$,
and $C$ are emu$\cdot$mol$^{-1}$, emu$\cdot$mol$^{-1}$K$^{-1}$, and
emu$\cdot$K$\cdot$mol$^{-1}$, respectively.}
\begin{ruledtabular}
\begin{tabular}{cccc}
Samples&$\chi_{0}$ ($\times 10^4$)&$\alpha$ ($\times 10^7$)&$C$\\
\hline
$x$=0 & 3.27 &3.6&0.0045\\
$x$=0.01 & 2.85 &3.0&0.007\\
$x$=0.02 & 2.45 &4.1& 0.0056\\
$x$=0.03 & 2.36 &4.3& 0.0038\\
$x$=0.04 & 2.71 &3.4& 0.0021\\
$x$=0.05 & 2.22 &3.6& 0.0027\\
$x$=0.06 & 2.00 &3.8& 0.0035\\
$x$=0.10 & 1.44 &3.5& 0.0098\\
\end{tabular}
\end{ruledtabular}
\end{table}

It is noted that the maximum $T_c$ (6.5 K) in
LaFe$_{1-x}$Ni$_{x}$AsO is merely one fourth of that in
LaFeAsO$_{1-x}$F$_{x}$\cite{Kamihara08}, and half of that in
LaFe$_{1-x}$Co$_{x}$AsO\cite{WangC}. The lowered $T_c$ in the
Co-doped system was discussed in terms of the relatively small
As-Fe-As angle\cite{WangC}, according to an empirical structural
rule for $T_c$ variations.\cite{Structure-Tc} However, the As-Fe-As
angle of LaFe$_{0.96}$Ni$_{0.04}$AsO is almost the same as that of
LaFe$_{0.925}$Co$_{0.075}$AsO. Therefore, the much lowered $T_c$ in
LaFe$_{1-x}$Ni$_{x}$AsO system should be caused by the reason other
than structural aspect.

Let us turn to examine the normal-state property to find the
possible clues. The normal-state resistivity strikingly exhibits a
semiconducting-like behavior above $T_c$, as shown in Fig. 2. At
first glance, the resistivity upturn at low temperatures might be
ascribed to Anderson localization owing to the Ni incorporation,
which might account for the lowered $T_c$. This scenario of
disorder-induced localization would lead to a more profound
resistivity upturn or higher $T^*$ (resistivity minimum temperature)
with the increase of Ni doping. However, the $\rho(T)$ curves in
Fig. 2 show that $T^*$ \emph{decreases} monotonically with
increasing $x$. Therefore, the evolution of the resistivity upturn
with Ni doping suggests that Anderson localization is unlikely to be
the main reason for the lowered $T_c$.

In the framework of a coherent-incoherent scenario\cite{DaiSi}, the
itinerant carriers and the local magnetic moments coexist in the
undoped iron arsenides. Based on the logarithmic upturn of
resistivity, a spin-flip scattering between the itinerant charge
carriers and the local moments in the undoped FeAs layers has very
recently been proposed.\cite{DaiCao}. It is noted that the spin-flip
scattering (actually analogue to Kondo effect) is already there in
the parent compound. Upon electron doping, $T^*$ is suppressed due
to the decrease of $N(E_F)$, as the Kondo energy scale
$T_{\text{K}}\propto \sqrt{JN(E_{F})}\text{exp}(-1/JN(E_{F}))$ with
$J$ being the Kondo coupling constant. In the case of Ni doping,
both extra itinerant 3$d$ electrons and \emph{stabilized} local
moments (as inferred from the band structure calculation\cite{Xu})
are introduced. For the same electron doping level, one would expect
an enhanced spin-flip scattering in the Ni-doped system. The
spin-flip scattering competes with the SC Cooper pairing, which
explains the suppression of $T_c$. Therefore, the narrow SC region
as well as the much lowered $T_c$ in LaFe$_{1-x}$Ni$_{x}$AsO system
is here ascribed to be a combined effect from the competing
Kondo-like interactions, Anderson localization, as well as the
structural variation.

Very recently, the effectiveness of Ni-doping for SC has been also
demonstrated in BaFe$_{2}$As$_{2}$ (Ref.~\cite{Li}) and CaFeAsF
(Ref.~\cite{Matsuishi}) systems. The $T_{\text{c,max}}$ are 20.5 K
and 12 K, respectively. The variations in $T_{\text{c}}$ are
possibly due to the Kondo-like interactions (not significant in
Ni-doped BaFe$_{2}$As$_{2}$) as well as the structural difference.
The bond angle of As-Fe-As in CaFeAsF is significantly smaller than
that of LaFeAsO.

Fig. 7 summarizes a SC phase diagram for LaFe$_{1-x}$Ni$_{x}$AsO
system. With Ni-doping, the SDW order is suppressed, followed by the
emergence of superconductivity. The SC region is particularly narrow
and the $T_c$ is remarkably low, as compared with those of
LaFeAsO$_{1-x}$F$_{x}$ and LaFe$_{1-x}$Co$_{x}$AsO (see Table III).
The normal state is divided by the line of $T^*$ into metallic and
semiconducting regions. The "optimal" doping occurs at
$x_{\text{opt}}$=0.04, which is about half of the $x_{\text{opt}}$
of LaFe$_{1-x}$Co$_{x}$AsO. This observation further demonstrates
the itinerant character of Ni 3$d$ electrons. Here we emphasize that
the occurrence of superconductivity by Ni doping contrasts sharply
with the cuprate superconductors, where the substitution of Cu with
Ni in CuO$_2$ planes severely destroys the
superconductivity\cite{Tarascon}.

\begin{figure}
\includegraphics[width=7cm]{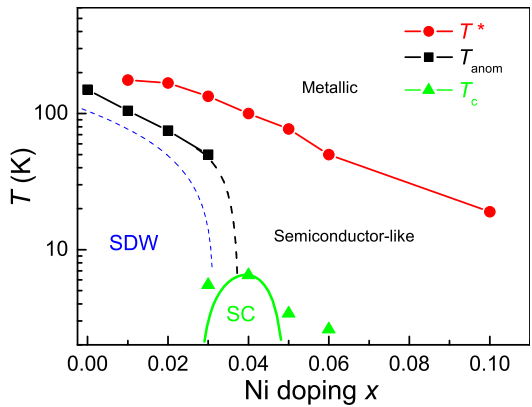}
\caption{(Color online) Electronic phase diagram for
LaFe$_{1-x}$Ni$_{x}$AsO, showing a narrow SC window. The dashed blue
line schematically shows the SDW transition temperature, which is
tens of kelvin below $T_{\text{anom}}$ (Ref.~\cite{Dai}). Note that
the vertical axis is in logarithmic scale.}
\end{figure}

\begin{table}
\caption{\label{tab:table3}Comparison of SC phase diagrams in
LaFe$_{1-x}$Co$_{x}$AsO\cite{WangC} and LaFe$_{1-x}$Ni$_{x}$AsO
(present work). $T_{\text{c,max}}$ denotes the maximum $T_c$ at
optimal doping level $x_{\text{opt}}$.}
\begin{ruledtabular}
\begin{tabular}{lcr}
System&LaFe$_{1-x}$Co$_{x}$AsO&LaFe$_{1-x}$Ni$_{x}$AsO\\
\hline
SDW region & 0 $\leq x < $ 0.025 & 0 $\leq x < $ 0.02\\
SC region & 0.025 $\leq x\leq $ 0.125 & 0.03 $\leq x\leq $ 0.06\\
$T_{\text{c,max}}$(K) & 13 & 6.5\\
$x_{\text{opt}}$ & $\sim$ 0.075 & $\sim$ 0.04\\
Normal-state $\rho$ & semiconducting & semiconducting\\
\end{tabular}
\end{ruledtabular}
\end{table}

\begin{acknowledgments}
We would like to thank Y. Liu, Q. Si, H.Q. Yuan and G.M. Zhang for
useful discussions. This work is supported by the NSF of China
(Contracts No. 10674119 and No. 10634030), National Basic Research
Program of China (Contract No. 2007CB925001) and the PCSIRT of the
Ministry of Education of China (Contract No. IRT0754) and RGC in
HKSAR.
\end{acknowledgments}


\begin{thebibliography}{00}
\bibitem{note}Here we ignore the discussion of Landau diamagnetism, because Landau susceptibility is 1/3 of $\chi_{\text{P}}$
when $m\approx m^*$.
\end{thebibliography}
\end{document}